%% file: main.tex
\documentclass[conference]{IEEEtran}
\IEEEoverridecommandlockouts
\usepackage{cite}
\usepackage{amsmath,amssymb,amsfonts}
\usepackage{algorithmic}
\usepackage{graphicx}
\usepackage{textcomp}
\usepackage{xcolor}
\newcommand{\Z}{\ensuremath{\mathbb{Z}}}
\newcommand{\B}{\ensuremath{\mathbb{B}}}
\newcommand{\M}{\ensuremath{\mathcal{M}}}
\newcommand{\ITE}[3]{\text{ITE}(#1, #2, #3)}

\usepackage{caption}
\usepackage{subcaption}
\usepackage{booktabs}

\usepackage[linesnumbered,ruled,noend]{algorithm2e}
\DontPrintSemicolon%
\SetKwFunction{SAT}{SAT}
\SetKwFunction{solve}{Solve}
\SetKwFunction{build}{Build}
\SetKwFunction{split}{Split}
\SetKwData{false}{false}\SetKwData{true}{true}
\SetKwInOut{Input}{input}
\SetKwInOut{Output}{output}
\SetKwInOut{Prec}{precondition}
\SetKw{Func}{Function}
\SetKw{Break}{break}

\newtheorem{theorem}{Theorem}[section]
\newtheorem{example}[theorem]{Example}

\definecolor{citeblue}{rgb}{0.1,0,.4}
\usepackage[pdftex%
,colorlinks=true%
,bookmarks=true%
,linkcolor=citeblue%
,citecolor=citeblue%
,urlcolor=blue%
,plainpages=false]{hyperref}
\begin{document}

\title{Towards Learning Infinite SMT Models\\
  (Work in Progress)
\thanks{%
The results were supported by the Ministry of Education, Youth and Sports
within the dedicated program ERC~CZ under the project \emph{POSTMAN}
no.~LL1902.
This article is part of the \emph{RICAIP} project that has received funding
from the European Union's Horizon~2020 research and innovation programme under
grant agreement No~857306.}}

\author{\IEEEauthorblockN{1\textsuperscript{st} Mikol\'{a}\v{s} Janota}
\IEEEauthorblockA{%
  \textit{Czech Technical University in Prague}\\
  Prague, Czech Republic\\
https://orcid.org/0000-0003-3487-784X}
\and
\IEEEauthorblockN{2\textsuperscript{nd} Bartosz Piotrowski}
\IEEEauthorblockA{%
\textit{Czech Technical University in Prague}\\
Prague, Czech Republic\\
https://orcid.org/0000-0002-1699-018X}
\and
\IEEEauthorblockN{3\textsuperscript{nd} Karel Chvalovsk\'y}
\IEEEauthorblockA{%
  \textit{Czech Technical University in Prague}\\
  Prague, Czech Republic\\
https://orcid.org/0000-0002-0541-3889}
}

\maketitle

\begin{abstract}
  \input{abstract}
\end{abstract}

\begin{IEEEkeywords}
SMT, infinite models, piecewise linear
\end{IEEEkeywords}

\section{Introduction}

Formulas with quantifiers remain a major challenge for
\emph{satisfiability modulo theories}~(SMT)~\cite{barrett2018satisfiability}
solvers. This is especially true when quantifiers are combined with
uninterpreted functions. Predominantly, SMT solvers tackle quantifiers by
gradually instantiating them by ground terms. Such terms may be chosen by a variety
of techniques, which themselves have further degrees of freedom that are
addressed by specific heuristics in the concrete implementations of the
solvers.

One of the oldest techniques is \emph{e-matching}~\cite{DetlefsNS05}, which is
mainly syntactic and does not guarantee any completeness. Refutation
completeness under certain conditions can be achieved by simple
\emph{enumerative instantiation}~\cite{ReynoldsBF18,janota2021fair}. Term
generation can also be further focused by syntactic
guidance~\cite{niemetz-tacas21} or conflicts~\cite{reynolds-fmcad14}.

A quantifier instantiation technique that stands out is \emph{model-based
quantifier instantiation (MBQI)}~\cite{ge-moura-cav09}, which unlike the
above-mentioned enables showing a formula SAT\@. This involves constructing a
sequence of candidate models that are checked against the formula and
drive further instantiations. In their earlier work, Bradley and Manna identify
the \emph{array property}~\cite{bradley-vmcai06,bradley-manna07}, for which
they show that a finite set of instantiations is always complete---MBQI can be
seen as an instantiation generalization of this approach~\cite{moura-ijcar10}.

An interesting question arises in the context of MBQI and that is how to
construct the candidate models based on the models of the current ground part
(which is gradually being strengthened by further instantiations). In this
paper, we propose to attempt to learn piecewise linear functions to represent
the candidate models. We use simple, fast algorithms that construct such functions
based on the values obtained from the ground model. In that sense, our approach
is purely semantic---it ignores the syntactic form of the formula, which
contrasts with the current techniques that are mainly syntactically
driven~\cite[Sec.~5]{moura-ijcar10}.

We implement the proposed techniques in the state-of-the-art SMT solver
cvc5~\cite{cvc5} and report encouraging results, where number of SAT responses
increases and does not incur a slowdown for UNSAT problems.

To the best of our knowledge such learning had not been attempted in the context of
MBQI\@. Some related approaches appear in the literature, finite models have
been learned in the context of finite model finding~\cite{janota-lpar18}.
Invariants and termination conditions in the context of software verification
have been constructed in an analogous
fashion~\cite{adje-vmcai15,urban-sas13,urban-sas14}. Another related
research direction is function and program \emph{synthesis}. Notably,
Barbosa~et~al.~\cite{barbosa-fmcad19} learn syntactically driven decision
trees to construct functions---while this is in the context of the SMT solver
cvc5~\cite{cvc5}, it is separate from the main solver so for instance it is not meant for
UNSAT problems. In recent work, Parsert~et~al.~show that state-of-the-art
synthesis approaches often fail when faced with SMT problems~\cite{lpar23}.

\section{Learning Models in MBQI}\label{s:algorithm}
MBQI works in iterations, where in each iteration there is a candidate model
$\M$ satisfying the ground part $\phi_g$. Let us assume that the formula being
solved is of the form $(\forall\mathbf{x}\,\phi)$ for some quantifier-free $\phi$
and a vector of variables $\mathbf{x}$. A sub-SMT call is issued to check
weather $\M$ satisfies $(\exists\mathbf{x}\,\lnot\phi)$. If it does not, the
formula is satisfied and the model $\M$ is also a model of the original
formula. If it is satisfied by some vector $\mathbf{c}$, then the ground
formula $\phi_g$ is strengthened by an instantiation of
$(\forall\mathbf{x}\,\phi)$ based on $\mathbf{c}$---for simplicity one may imagine that
$\mathbf{c}$ is plugged into $\mathbf{x}$. Hence, the process either
terminates by finding a model of the original formula, or when the ground part
becomes unsatisfiable (meaning that the original formula is also
unsatisfiable), or goes on indefinitely. This contrasts with other quantifier
instantiation techniques that are only able to prove unsatisfiability.

\begin{example}\label{example:mbqi}
  The following is a possible (non-terminating) run of MBQI on the formula
  $(\forall x:\Z\,f(x)>x)$. ITE abbreviation below signifies an
  \textit{if-then-else} expression.
 \[
   \begin{array}{llll}
     \text{iteration} & \phi_g             & f(x)            & c \\\toprule
     0           & \text{true}        & 0               & 0 \\
     1           & f(0)>0             & 1               & 1 \\
     2           & f(0)>0\land f(1)>1 & \ITE{x=0}{1}{2} & 2 \\
     \vdots      & \vdots             & \vdots          & \vdots \\
   \end{array}
 \]

\end{example}

Example~\ref{example:mbqi} shows that for satisfiable formulas MBQI can easily
diverge, even though it could terminate quickly if it guessed a different
candidate model, for instance $f(x)=x+1$. Neither Z3 nor cvc5
solves this formula. In this work, we focus specifically on simple model
candidates that can be composed of linear functions.

Here we focus on problems from linear integer arithmetic with uninterpreted
functions (UFLIA). In each iteration of the MBQI loop, we assume that we are
given for each $n$-ary function or predicate $f$ finitely many points
$(\mathbf{a}, v)$, where $\mathbf{a}$ is a vector of $n$ values for the
arguments and $v$ is the value of $f$ for these arguments, i.e.,
$f(\mathbf{a})=v$. The value is either an integer (for functions) or Boolean (for predicates).
These function points appear naturally in MBQI as a model of the ground part. So
for instance, in iteration 2 of Example~\ref{example:mbqi}, the model for the
ground part might assign $f(0)=2$ and $f(1)=3$, allowing us to propose
$f(x)=x+2$.

Currently, we consider each function or predicate in the formula separately.
Hence, the objective is fitting a piecewise linear function to these points,
which is essentially a synthesis task. Here we also need to take into
account that the points are in the space of integers and they must be matched
perfectly, not just approximately as in some methods used in statistical
learning, \textit{cf.}~\cite{landwehr2005}. Further, the process needs to be efficient since
it may be invoked many times during individual iterations of MBQI\@.

\input{alg_greedy}

A straightforward approach to synthesizing a piecewise linear function is a
greedy one, where we first sort the points according to some criterion (e.g.,
lexicographically) and then try to greedily connect adjacent points into a
single hyperplane. This approach is outlined in Algorithm~\ref{alg:greedy}. The
points are organized in a list and to check if the current point fits
onto the hyperplane under construction, we add a corresponding constraint to the set
of equations $C$.
These equations are of the form $\mathbf{a}^T\mathbf{y}+c=v$,
where $\mathbf{a}\in\Z^n,c\in\Z$ and $\mathbf{y}$  is a vector of integer
variables. These are linear
Diophantine equations solvable in polynomial time~\cite[pgs.~343--345]{knuth98}~\cite{griggio12}.

Once the set of constraints $C$ becomes unsatisfiable, a new segment
(hyperplane) needs to be started. In order to construct an SMT term, we use an
if-then-else (ITE) expression. The splitting condition must be such that the
points already covered and the points yet to be covered become disjoint under
this condition. How exactly this split is done is represented by the function
$\split$ in the pseudocode. In our implementation, we use a
lexicographic order on $P$, which lets us also easily split the points as
follows. If the last covered point is $a_0, a_1, \dots$ and the point yet to be
covered is $a'_0, a'_1, \dots$, the condition is
\[x_0<a'_0 \lor (x_0=a'_0\land x_1<a'_1) \lor \dots\]
Additionally, this condition is simplified so that we consider $x_i$ only if
$a_j=a'_j$ for $j<i$. If all the given points already fit on the hyperplane
under construction, no splitting is needed.

\input{alg_pred_smarter}

To treat predicates, rather than equations, as a primitive we use inequalities
of the form $\mathbf{s}^T\mathbf{a}\geq c$, $\mathbf{s}\in\Z^n,c\in\Z$. An
analogous greedy algorithm could be used to split a list of points into
segments. However, only very simple predicates can be learned by this approach,
since each segment is only able to separate points by a single hyperplane. So
for instance equality cannot be learned.

An alternative is to use decision trees but here the question is what should be
the predicates used in the internal nodes of the tree? In statistical machine
learning, decision trees split on the value of a single feature (variable) and
this seems to be too limiting since this enables capturing only limited
interactions between variables---again, equality would not be
learnable. Hence, we apply a hybrid approach where the points are split into two
parts by \emph{some} hyperplane and the rest is classified recursively. Effectively, we
are building a decision tree where each branch corresponds to a convex
polyhedron and each polyhedron should only contain points of one value.

This approach is outlined in Algorithm~\ref{alg:predSmarter}, which
looks greedily for a hyperplane splitting the encountered points into positive
and negative. Since this hyperplane might not split the points perfectly, each
``half'' is further refined recursively.

For the Algorithm~\ref{alg:predSmarter} to terminate, the chosen hyperplane
must split at least one positive and one negative point.  Further, the algorithm
is sensitive to the order in which the points are added to the constraints~$C$.
We use a simple heuristic for this order. We first focus on a pair of points
with different values (one false, one true). Since there may be many such pairs,
we first sort by lexicographic order and pick a pair of adjacent points $(p_i,p_{i+1})$ with
different values. There still may be multiple such pairs
and we pick such pair that maximizes information gain~\cite{quinlan86} by
looking at the points left and right of the pair---i.e., by looking at the
subsets $\{p_1,\dots,p_i\}$ and $\{p_{i+1},\dots,p_n\}$, as if they were split by the
hyperplane currently constructed (even though this might not eventually be
true, since we only guarantee that $p_i$ and $p_{i+1}$ are split). Other points are
added into the constraints by going first right and then going left from the
splitting pair---this order was chosen arbitrarily.

Compared to Algorithm~\ref{alg:predSmarter}, the current implementation in fact
stops the greedy search on the first UNSAT response from the sub-solver. The
reason is that we want to avoid inefficiencies in the sub-solver, which is
currently used in an incremental setting without push and pops.

\begin{figure*}
\begin{subfigure}{.245\textwidth}
\includegraphics[width=1.0\textwidth]{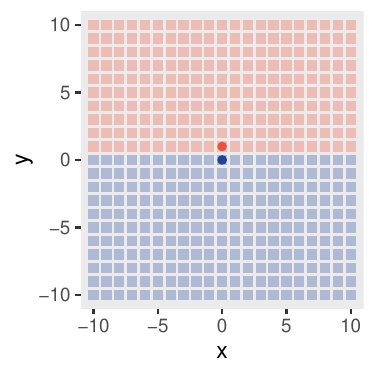}
\end{subfigure}
\begin{subfigure}{.245\textwidth}
\includegraphics[width=1.0\textwidth]{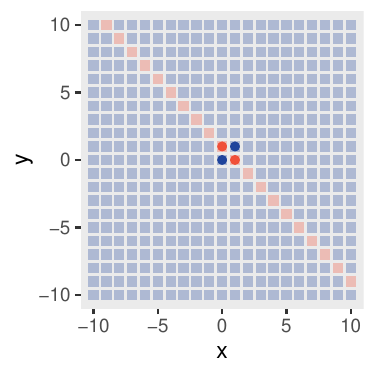}
\end{subfigure}
\begin{subfigure}{.245\textwidth}
\includegraphics[width=1.0\textwidth]{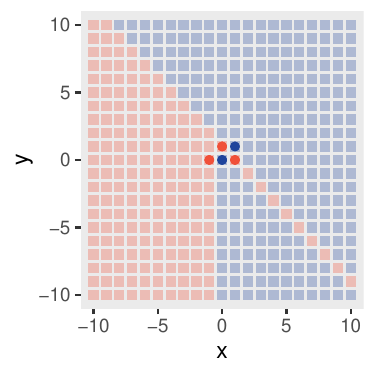}
\end{subfigure}
\begin{subfigure}{.245\textwidth}
\includegraphics[width=1.0\textwidth]{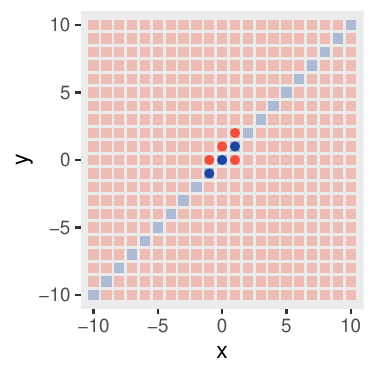}
\end{subfigure}
\caption{Four snapshots of the iterative process of finding an infinite model for a binary relation $R$, as described in Algorithm~\ref{alg:predSmarter}.
The relation $R$ is constrained by two assertions:
$R(x,y) \Rightarrow x = y$
and
$x = y \Rightarrow R(x, y)$.
Yellow color signifies \textit{false} and blue color signifies \textit{true}.
Dots in the center are points with boolean values assigned by the solver.
Rectangles in the background depict the current relation assigned to $R$.
The last figure shows the state in which the algorithm found the desired
relation.
}%
\label{fig:name}
\end{figure*}

Fig.~\ref{fig:name} visualizes how equality can be found using this algorithm.
The example shows the evolution of models for the formula
$(\forall xy:\Z\,R(x,y)\Rightarrow x=y)\land(\forall xy:\Z\,x=y\Rightarrow
R(x,y))$, which unambiguously defines $R$ as the equality.
Also, this instance is not solved by neither Z3 nor cvc5.

In the final iteration of the MBQI loop, the learning algorithm is given the positive
points $\{(0,0), (1,1), (-1,-1)\}$ and the negative points $\{(-1,0), (0,1),
(1,0), (1,2)\}$. The recursive algorithm learns the term $\ITE{x-y\geq
0}{-x+y\geq 0}{\text{false}}$, which  can be seen as the intersection of $x\geq
y$ and $y\geq x$, as expected.

\section{Experiments}\label{sec:experiments}
The proposed algorithms were implemented in cvc5~\cite{cvc5} in the main branch.
The experiments were run with the time limit of 30\,s. We evaluate on two sets
of benchmarks. The first set was obtained by taking SMT-LIB benchmarks from
UFLIA that do not contain any uninterpreted sorts; these tend to be
unsatisfiable. The second set is obtained by taking fragments of UFLIA
benchmarks---we describe the process in the following subsection.
The solver cvc5 is run with e-matching turned off, i.e.\  MBQI only.
We compare two versions: one where predicates are synthesized by
Algorithm~\ref{alg:greedy} (\emph{non-smart}) and Algorithm~\ref{alg:predSmarter}
(\emph{smart}).

The results for these two sets of problems are summarized in
Tables~\ref{tab:unsat-results} and~\ref{tab:sat-results}, respectively. The
results indicate that on satisfiable instances, our semantic-guided syntactic
outperforms the syntactic approach of Z3. Our implementation does
not incur any slowdown on the unsatisfiable instances. In fact, a handful of
instances are solved on top of the default MBQI\@.

\subsection{Generation of satisfiable problems}\label{sec:generation}
We consider a simple technique to generate interesting satisfiable fragments
from a given SMT formula---understood as a conjunction of assertions. The idea
is to consider some parameter $k$ and extract fragments that contain $k$
uninterpreted functions. Given an SMT formula, we choose $k$ uninterpreted
function symbols that appear in the formula and filter out the conjuncts that
are only weakly related to these $k$ symbols. Let us describe the process for
$k=2$. Consider an SMT formula $\phi$ and two uninterpreted function symbols
$f$ and $g$ that occur in $\phi$. Consider subformula $\psi$ in $\phi$ (one of
the conjuncts) of the form $(\forall\mathbf{x}\,\psi')$, where $\mathbf{x}$
is an arbitrary set of variables. We will say
that $\psi$ is \emph{in the $f,g$ fragment of $\phi$} if it contains at least
one $f$ or $g$ and no other uninterpreted functions; there is no limitation on
constants. \emph{The global $f,g$ fragment of $\phi$} is defined as the
conjunction of all the $f,g$ fragments in $\phi$. We remark that in the current
implementation, we only consider subformulas that are denoted by the users as
separate assertions---this could be relaxed.

Since the current implementation only supports generation of functions and
predicates on integers, we only consider UFLIA problems and replace all of
interpreted sorts by \texttt{Int}.

\begin{table}
\centering
\begin{tabular}{lc}
\toprule
solver & solved: UNSAT \\
\midrule
standard MBQI       & 3831\\
ours non-smart MBQI & 3831\\
ours smart MBQI     & 3835\\
Z3                  & 5809\\
\bottomrule
\end{tabular}
\caption{Problems solved within 30\,s time limit by four solvers in a benchmark
being a set of these UFLIA problems that does not have sort declarations. There
were no SAT results in this benchmark. The total number of problems is 6288.}%
\label{tab:unsat-results}
\end{table}

\begin{table}
\centering
\begin{tabular}{lccc}
\toprule
solver & solved: SAT & solved: UNSAT & solved: total \\
\midrule
standard MBQI       & 18843 & 7863 & 26706 \\
ours non-smart MBQI & 29456 & 7863 & 37319 \\
ours smart MBQI     & 31977 & 7863 & 39840 \\
Z3                  & 28380 & 7482 & 35862 \\
\bottomrule
\end{tabular}
\caption{Problems solved within 30\,s time limit by four solvers in a benchmark
based on UFLIA problems modified to make most of the problems SAT by removing
some assertions, and where all declared sorts were substituted with
\texttt{Int}. The total number of problems is 69692.}%
\label{tab:sat-results}
\end{table}

\section{Summary}\label{s:future}
In this short paper we propose algorithms for the construction of
piecewise-linear model candidates in the context of model-based quantifier
instantiation (MBQI), which is a powerful instantiation technique for solving
SMT problems with quantification. In essence, this means learning an
\emph{infinite} model based on \emph{finite} information. The experimental
evaluation shows that many new satisfiable problems can be solved by the
proposed approach and at the same time it does not slow down the solver for
unsatisfiable problems.
In the future, we would like to explore closer collaboration with the
sub-solvers and the main algorithm, \textit{cf.}~\cite{abate-jar23}.

\subsection*{Acknowledgments}
We would like to thank Chad Brown for numerous
discussions on the topic and Andrew Reynolds for giving useful pointers
regarding the codebase of cvc5.

\bibliographystyle{IEEEtran}
\bibliography{refs}
\end{document}

%% file: abstract.tex
This short paper proposes to learn models of satisfiability modulo theories
(SMT) formulas during solving. Specifically, we focus on infinite models for
problems in the logic of linear arithmetic with uninterpreted functions (UFLIA).
The constructed models are piecewise linear. Such models are useful for
satisfiable problems but also provide an alternative driver for model-based
quantifier instantiation (MBQI).

%% file: alg_greedy.tex
\begin{algorithm}[t]
  \Func \build($P$)\;
    \Input{list of function points $P\subset\Z^n\times\Z$}
    \Output{function $\Z^n\mapsto\Z$}
    \BlankLine
    $C\gets\{\}$\tcp*[r]{set of constraints}
    $S\gets\{\}$\tcp*[r]{set of covered points}
    \While{$P\neq\emptyset$}{
      $\mathbf{a},v\gets\text{head}(P)$\;
      $C'\gets C\cup\{\mathbf{a}^T\mathbf{y} + c = v\}$\;
      \If{$\text{SAT}(C')$}{
        $C\gets C'$\;
        $P\gets\text{tail}(P)$\;
        $S\gets S\cup\{(\mathbf{a},v)\}$
      } \Else{
        \Break
      }
    }
    $\mathbf{s}, c\gets\solve(C)$\;
    $\text{segment}\gets\lambda\mathbf{x}.\,\mathbf{s}^T\mathbf{x}+c$\;
    \lIf{$P=\emptyset$}{
      \Return\text{segment}
    } \lElse {
    \Return $\ITE{\split(S, P)}{\text{segment}}{\build(P)}$
    }
 \caption{Greedy function construction}\label{alg:greedy}
\end{algorithm}

%% file: alg_pred_smarter.tex
\begin{algorithm}[t]
  \Func \build($P$)\;
    \Input{positive and negative points $P\subset\Z^n\times\B$}
    \Output{predicate $\Z^n\mapsto\B$}
    \BlankLine
    \lIf{$P$ all negative}{\Return $\lambda \mathbf{x}.\text{false}$}
    \lIf{$P$ all positive}{\Return $\lambda \mathbf{x}.\text{true}$}
    $C\gets\{\}$\tcp*[r]{set of constraints}
    \ForEach{$(\mathbf{a},b)\in P$}{
      \lIf{b}{
        $C'\gets C\cup\{\mathbf{a}^T\mathbf{y} \geq c\}$
      } \lElse {
        $C'\gets C\cup\{\mathbf{a}^T\mathbf{y} < c\}$
      }
      \lIf{$\text{SAT}(C')$}{
        $C\gets C'$
      }
    }
    $\mathbf{s}, c\gets\solve(C)$\;
    $P^+\gets\build(\{(\mathbf{a},b)\in P \mid \mathbf{s}^T\mathbf{a}\geq c\}$)\;
    $P^-\gets\build(\{(\mathbf{a},b)\in P \mid \mathbf{s}^T\mathbf{a}< c\}$)\;
    \Return $\lambda\mathbf{x}.\,\ITE{\mathbf{s}^T\mathbf{x}\geq c}{P^+}{P^-}$
    \caption{Recursive predicate splitting}\label{alg:predSmarter}
\end{algorithm}